\documentclass[12pt,final]{revtex4}
\usepackage{psfrag}
\usepackage{amssymb}
\usepackage{textcomp}
\usepackage{units}
\usepackage{graphicx}
\usepackage{natbib}

\begin{document}

\title[Synchronization of organ pipes]{Synchronization of organ pipes: 
experimental observations and modeling}

\author{M. Abel$^{1,2}$}
%  {address={Institute of Physics, University of Potsdam}}

%\affiliation{Institute of Physics, University of Potsdam, 14415 Potsdam,  Germany}
%\affiliation{UP Transfer GmbH at the  University of  Potsdam, 14469 Potsdam, Germany}
\author{S. Bergweiler$^{2}$}
%  {address={Institute of Physics, University of Potsdam}}
%\affiliation{Institute of Physics, University of Potsdam,\\14415 Potsdam, Germany}

\author{R. Gerhard-Multhaupt$^1$}
%  {address={Institute of Physics, University of Potsdam}}

\affiliation{$^1$ Institute of Physics, University of Potsdam,14415 Potsdam, Germany}
\affiliation{$^2$ UP Transfer GmbH at the  University of  Potsdam, 14469 Potsdam,
  Germany}
%\classification{43.75.+a,43.28.+h}
%\received{\today}
\date{\today}

\begin{abstract}
We report measurements on the synchronization properties of organ
pipes. First, we investigate influence of an external acoustical
signal from a loudspeaker on the sound of an organ pipe. Second, the
mutual influence of two pipes with different pitch is analyzed.  In
analogy to the externally driven, or mutually coupled self-sustained
oscillators, one observes a frequency locking, which can be
explained by synchronization theory. Further, we measure the
dependence of the frequency of the signals emitted by two mutually
detuned pipes with varying distance between the pipes.  The spectrum
shows a broad ``hump'' structure, not found for coupled oscillators.
This indicates a complex coupling of the two organ pipes leading to
nonlinear beat phenomena.

\vspace{2ex} \noindent PACS numbers: 05.45.Xt, 47.52.+j, 43.75.+a, 43.28.+h
\end{abstract}

%\pacs{43.75.+a, 05.45.Xt, 43.28.+h, 47.52.+j}
%43.75.+a Music and musical instruments
%43.28.+h Aeroacoustics and atmospheric sound
%47.52.+j Chaos (see also 05.45.-a Nonlinear dynamics and nonlinear dynamical systems; 83.60.Wc Flow instabilities)
%05.45.Xt Synchronization; coupled oscillators

\maketitle \hspace{4ex}

\section{Introduction}

Sound production in organ pipes is traditionally described as a
generator-resonator coupling. In the last decades, research has been
concerned with the complex aeroacoustic processes which lead to a
better understanding of the sound generation in a flue organ pipe.
The process of sounding a flue-type organ pipe employs an airstream
directed at an edge, the labium of an organ pipe. An oscillating
``air sheet'' is used to describe the situation in which the
oscillations of the jet exiting from the flue are responsible for
the creation of the pipe sound \cite{Fabre-00,RossingFletcher-97}.
Using the ``air sheet'' terminology, it is pointed out that the
oscillation is controlled not by pressure, as in earlier
investigations
\cite{Cremer-Ising-68,Coltman-68,Coltman-92ab,Fletcher-93,Nolle-83},
but by the air flow
\cite{Coltman-76,Verge-94,Verge-97a,Verge-97b,Fabre-Hirschberg-96,Segoufin-04}.

The situation becomes more involved if two flue organ pipes are
close in sounding frequency and in spatial distance. Then a
synchronization of the pipes, a frequency locking, occurs
\cite{Rayleigh2-45,Bouasse-29,Stanzial-00,Angster-93}. This has a
direct importance for the arrangement of organ pipes in a common
Orgelwerk.  The effect has been known by organ builders for a long
time and taken into account intuitively in the design of organs
\cite{Schuke-private}. Some measurements of the acoustic field or of
its dependence on the mutual coupling or on the distance between the
pipes have been reported first in the early 20th century
\cite{Bouasse-29}, but no theoretical explanation was given. In this
article, we report acoustic measurements and give a possible
explanation by means of modern synchronization theory
\cite{Pikovsky-Rosenblum-Kurths-01}. To a certain approximation, an
organ pipe can be considered as a self-sustained oscillator,
explained in detail below. In this context, the synchronization of a
pipe by an external, acoustical  force is important, such that we
carried out additional measurements, where a pipe stands aside a
loudspeaker whose frequency is tuned around the pipes pitch. As well
in this case, the pipe is found to be synchronized.

As a side result of these measurements, we can clarify a discussion
about the nature of the strong amplitude decrease for two coupled
pipes, already observed by Lord Rayleigh \cite{Rayleigh2-45} and
later by very detailed measurements of Bouasse \cite{Bouasse-29}. As
a possible interpretation, the so-called oscillation death has been
given in \cite{Pikovsky-Rosenblum-Kurths-01}, which means that there
is an oscillation breakdown at the the pipe mouth, and all the
energy is dissipated.  From our results, we rule out such a scenario
and suggest an antiphase oscillation which yields destructive
interference of the emitted acoustic waves. This holds for two
neighboring pipes, as well as for the pipe standing aside a
loudspeaker.

This article is structured as follows: In section \ref{sec:basics}
we explain briefly the generation of sound in an organ pipe and
review basic concepts from synchronization theory.  In section
\ref{sec:experiment} we report on detailed measurements of the
synchronization of a loudspeaker positioned directly on the side of
a pipe and observe two adjacent pipes. The results are interpreted
within the frame of synchronization theory. We give an explanation
of why a model of two oscillators works out so well, as indicated in
\cite{Stanzial-00}. Further, the dependence of the frequency
spectrum on the distance between two detuned pipes is investigated
for a fixed amount of detuning. Finally, we conclude with section
\ref{sec:Conclusion}.

\section{Basic principles}
\label{sec:basics}
\subsection{Sound generation in organ pipes}
\label{sec:sound} Sound generation in organ pipes has been
repeatedly investigated
\cite{Coltman-76,Fabre-Hirschberg-96,Rayleigh-82}. Here, the beauty
of musical sound generation is paired with complex aerodynamical
phenomena; their coupling to the acoustic field has been understood
to a reasonable degree in the last 30 years (see the review
\cite{Fabre-00}).

Let us consider a single organ pipe: The wind system blows with
constant pressure producing a jet exiting from the pipe flue.
Typical Reynolds numbers, corresponding to a free jet are of the
order of $10^3$, depending on the pressure supplied and the pipe
dimensions, see \cite{Fabre-Hirschberg-96,Pitsch-Angster}.
%This is a transitional regime roughly analogous to the scenario found behind a long cylinder
%\cite{Fabre-Hirschberg-96,Williamson-96}.
The jet exiting from the flue generates a pressure perturbation at the labium
of the pipe which travels inside the pipe resonator and is reflected at the
end of the resonator.  This pressure wave returns after time $T$ to the labium
where it in turn triggers a change of the phase of the jet oscillations.
After a few transients, a stable oscillation of an ``air sheet'' at the pipe
mouth is established. This oscillation of the wind field couples to the
acoustic modes and a sound wave is emitted. Thus, the system can be considered
as a damped linear oscillator (the resonator) which controls the periodic
energy supply of the air-jet by the air column oscillation.  We want to adapt
this idea of an organ pipe as a self-sustained oscillator
\cite{Fletcher-79,Fletcher-99} to understand the mechanism of mode- or
frequency locking \cite{Fletcher-78}, or synchronization.

\begin{figure}[h]
\includegraphics[width=0.4\textwidth]{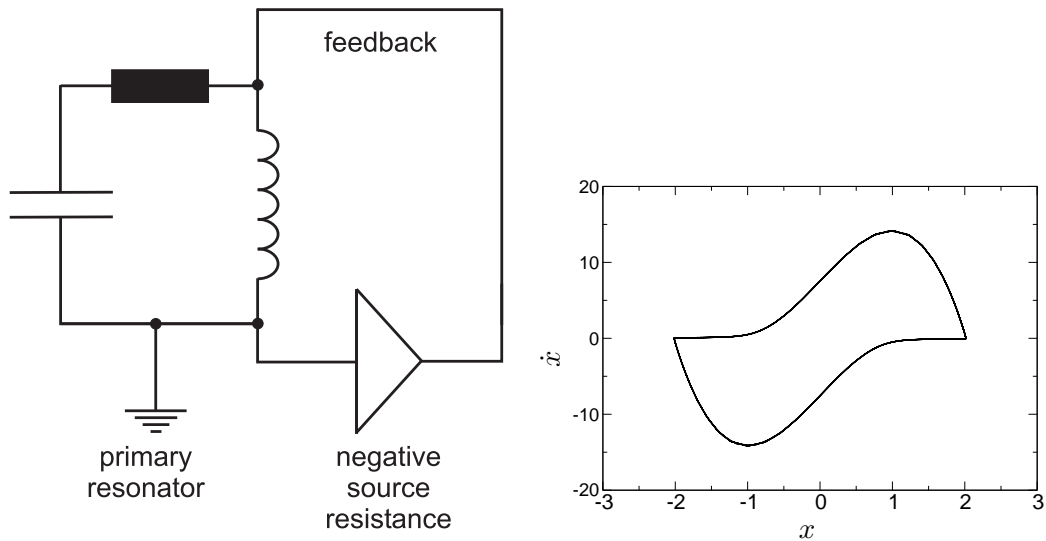}
\hspace{2ex}
\psfrag{Yaxis}{\small$\dot{x}$}
\psfrag{Xaxis}{\small$x$ }
\includegraphics[width=0.4\textwidth]{fig2.eps}
\vspace{2ex}\\
\caption{Left:Schematic diagram for a self sustained oscillator. Right:
  resulting limit cycle behavior.  }
\label{fig:vdP}
\end{figure}

One feature of a self sustained oscillator is the occurence of oscillations
even for constant driving. The physical mechanism is the balance of energy
losses and energy input. The classical example is the van der Pol oscillator
(Fig.~\ref{fig:vdP}), where a Triode (or more modern, a tunnel diode) acts as
``negative resistance'' whose response is fed back to an electrical oscillator.
In the case of an organ pipe, energy is supplied by the wind system, losses
are due to internal dissipation in the spatially extended resonator, and
radiation.  The negative resistance is represented by the air stream at the
pipe mouth and the feedback coupling is given by the boundary conditions
between the ``air sheet'' wind field and the air-column oscillations in the
resonator. It can be described by impedances \cite{Fletcher-76}, detailed
results for two coupled Helmholtz resonators are given in
\cite{Johansson-Kleiner-01}. Of course, the exact form of the limit cycle and
thus the acoustic oscillations depend on the details of energy losses, as the
quality factor of the resonators, the radiation of sound, and the
understanding of the energy supply by the coupling to the jet. The general
mechanism, however, remains untouched, since only a nonlinear growth of losses
and supply is needed for a self-sustained oscillator to work. Here, we do not
want to investigate these details, but rather emphasize the general character
of our investigations.

The sound emitted by this complex system can be evaluated by the
coupling of the flow and the acoustic modes in the framework of
aeroacoustic modeling on the basis of Lighthills analogy.  According
to Howe \cite{Howe-75,Howe-03,Howe-98}, the sound generation is
dominated by the singularity at the edge of the labium.  Numerical
approaches still suffer from the very expensive runs needed to
resolve the large range of excited scales and the proper choice of
boundary conditions \cite{Lele-97,Tam-97}.

\subsection{Synchronization of self-sustained oscillators}
\label{sec:synchro} In his book, Lord Rayleigh states ``When two
organ pipes of the same pitch stand side by side ... it may still go
so far as to cause the pipes to speak in absolute unison in spite of
inevitable small differences'' \cite{Rayleigh2-45}. He describes the
so-called ``Mitnahme-Effekt'' (loosely the take-along effect), known
by organ builders \cite{Stanzial-00,Angster-93}.  In
\cite{Pikovsky-Rosenblum-Kurths-01} the phenomenon is interpreted in
the frame of synchronization theory. Hereafter we will use the terms
frequency locking or synchronization to be synonymous with the
``Mitnahme-Effekt''.  In \cite{Angster-93}, the effect has been
analyzed heuristically, while Stanzial has investigated the
dependence of the frequency locking on the detuning or frequency
difference of two single pipes in \cite{Stanzial-00} by acoustic
measurement.  He modeled the effect by two coupled oscillators,
without giving a physical reason of the coupling. A treatment of the
mode-locking phenomenon, specific for musical instruments is very
nicely given in \cite{Fletcher-99}. In the following we will rely on
synchronization theory to show that {\it any} self-sustained coupled
oscillator generically shows synchronization.

Probably the most important feature of self-sustainment is the occurrence of an
attracting limit cycle. It appears due to two properties: nonlinearity, and
the energy balance between losses and driving. For a linear, damped oscillator
a limit cycle solution does not exist, the only possible attractor is the
trivial solution. Nonlinearity allows for the dependence of frequency on
amplitude which constitutes a mechanism to drive the system towards an
amplitude at which the regular oscillations are established. Since the
amplitude corresponds directly to the mechanical energy of the system, this is
right at the point of equality of losses and supply -- the limit cycle.

We want to sketch the principles for synchronization of an
oscillator with an external driving and then explain the interaction
of two oscillators. In the following we will discuss some equations
in terms of the angular frequencies and return later to frequencies
when the measurements are concerned.\\[-1.25ex]

%\subsubsection{Single Oscillator}
\underline{\it A Single Oscillator:} Let us consider the
equation for a harmonic oscillator with negative, nonlinear
resistance:
\begin{equation}
\ddot{x}= -\omega_0^2 x - \frac{d}{dt} f(x) = -\omega_0^2 x -
\dot{x} \frac{d f}{dx}  \; ,
\end{equation}
where $\omega_0$ is the angular frequency of the harmonic system and $f(x)$ is
a nonlinear function with at least one part of positive and another part of
negative slope. Now we discuss the basic principles by the concrete example of
the van der Pol oscillator and give general results without further
derivation, for details see \cite{Pikovsky-Rosenblum-Kurths-01}. In the case
of the van der Pol oscillator $f(x)=\alpha x - \frac{1}{3} \beta x^3$, and
$\omega_0^2= (LC)^{-1}$. The energy supply is accounted for by $\alpha$, the
losses by $\beta$.  If the inharmonicity is not too great, the amplitude on
the limit cycle differs little from the harmonic oscillator and the phase
coincides approximately with the harmonic one.  To get rid of fast
oscillations, one applies the method of averaging
\cite{Bogoliubov-Mitropolsky-61}: One inserts for the amplitude $x(t)=
A(t)\sin(\Theta_0)= A(t) \sin(\omega_0 t)$ and collects terms of the different
time scales 1 and $\omega_0$. So, one obtains an equation for the slowly
varying amplitude $A(t)$, and for the phase $\Theta_0$ here for the van der
Pol oscillator.
\begin{eqnarray}
\dot{A} &=& 1/2 \left(\alpha - \frac{1}{4} \beta A^2 \right) A\;,\\ \label{eq:SVE11}
\dot{\Theta}_0 &=& \omega_0\;.\label{eq:SVE12}
\label{eq:SVE1}
\end{eqnarray}
Eq.~\ref{eq:SVE11} describes the slow relaxation of the amplitude towards the
steady state. A limit cycle exists for $|A|^2 = \frac{4 \alpha}{ \beta} $.
The solution of the phase equation allows for the addition of an arbitrary
constant phase. Thus, one can shift the state along the cycle without changing
the energy. This allows for the adjustment of the phase, when the system is
driven, or coupled to another oscillator. This adjustment is exactly what we
understand as synchronization, or mode-locking.\\[-1.25ex]

\begin{figure}[h]
\includegraphics[angle=0,width=0.7\textwidth]{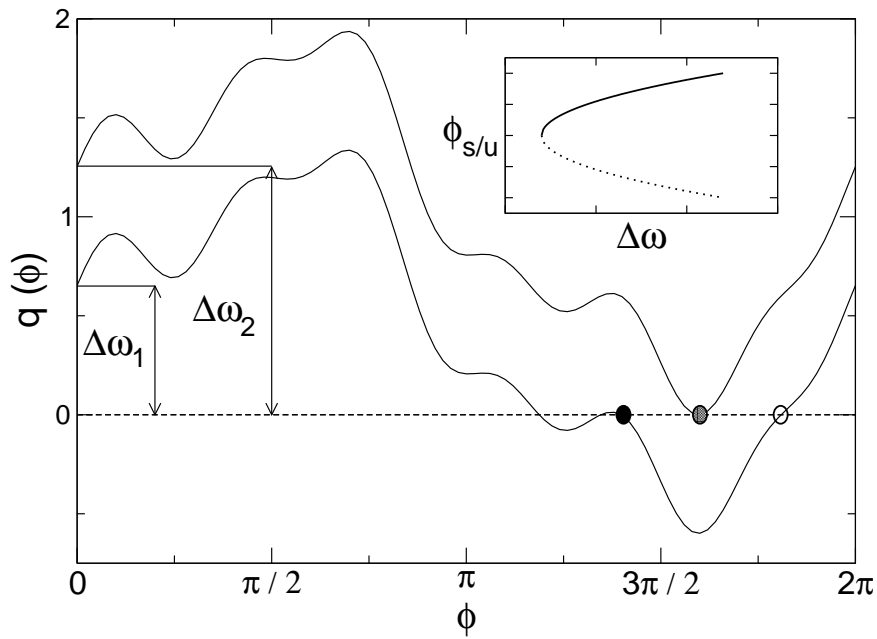}
\caption{Stable (filled circle) and unstable (open circle) fixed point
  $\Theta_{s}$, $\Theta_u$ for the Adler equation to illustrate the mode
  locking. If detuning or coupling is varied, the phase is adjusted
  accordingly by the system to the stable fixed point. For $\Delta
  \omega-\epsilon=0$ both fixed points merge; this results in a saddle-node
  bifurcation, as sketched in the inset; the straight/dotted line corresponds
  to the stable/unstable fixed point.}
\label{fig:Adler}
\end{figure}

%\subsubsection{Driven, Single Oscillator}
\underline{\it A Driven, Single Oscillator:}
If the system is driven externally by a harmonic force with angular frequency
$\omega_1\simeq\omega_0$, two time scales are present in the system, a fast one
$t_f=\frac{2\pi}{\omega_1}\simeq=\frac{2\pi}{\omega_0}$, and a slow one
  $t_s=\frac{2\pi}{\omega_1-\omega_0}$, and
  $t_s\ll t_f$.  The corresponding  equation
\begin{equation}
\ddot{x} +\omega_0^2 x - \dot{x}  \frac{d f}{dx} = \omega_1^2 R \cos(\omega_1 t) \; ,
\end{equation}
We want to investigate now the dependence on the slow time and average over
the fast one. To do so, we use the ansatz $x(t)= A(t) \sin(\omega_1 t +
\phi)$, with $\phi$ the slow phase (It is a bit easier working with a complex
amplitude, substituting $x=A e^{i\omega_1 t + \phi},\; \dot{x}=A\omega_1
e^{i\omega_1 t + \phi}$). By using again the averaging method one obtains
after a few straightforward steps \cite{Landa-96,Pikovsky-Rosenblum-Kurths-01}
%. To the first order Eq.~\ref{eq:SVE1} is
%recovered.
%Collecting the terms in $\omega_1$,  we find
%\begin{equation}
%2\omega_1 \dot{A} \cos(\omega_1 t) - A (\omega_1^2-\omega_0^2)  \sin(\omega_1 t)
%- \alpha A\omega_1 \cos(\omega_1 t)
%+ \beta A^3 \omega_1 \cos(\omega_1 t) \sin(\omega_1 t)
%=  \omega_1^2 R \cos(\omega_1 t)\;.
%\end{equation}
%Since now the phase is no longer free, the slow time is introduced in the
%phase by $A=a e^{i\phi}$, $a=|A|$; now two equations for phase and
%amplitude are obtained:
\begin{eqnarray}
\dot{a} &=& \frac{1}{2}\left(\alpha- \frac{1}{4}\beta a^2 \right)a -
\frac{\omega_1 R}{2} \cos \phi \;,\\
\dot{\phi} &=& \frac{1}{2} \frac{\omega_0^2-\omega_1^2}{\omega_1} +
\frac{\omega_1}{2}  \frac{R}{a} \sin \phi \;.
\end{eqnarray}
The second equation is known as the Adler equation
\cite{Adler-46,Pikovsky-Rosenblum-Kurths-01}; the general form for
the phase equation depends on the nonlinear function $f(x)$ and can
be seen as the difference of the equations for the two fast
variables $\omega_1$, and $\omega_0$:
\begin{equation}
\begin{array}{lcl}
\dot{\Theta}_1 &=& \omega_1\;,\\
\dot{\Theta}_2 &=& \omega_0+\epsilon q(\omega_0-\omega_1)\;,
\end{array}
\hspace{4ex}\dot{\phi} = \dot{\Theta}_2-\dot{\Theta}_1=\Delta \omega +
\epsilon q(\phi)
\label{eq:Adler}
\end{equation}
with a $2\pi$-periodic function $q$, and the two parameters detuning $\Delta
\omega = \frac{1}{2} \frac{\omega_0^2-\omega_1^2}{\omega_1} \simeq \omega_0 -
\omega_1 $ and locking term $\epsilon=\frac{\omega_1}{2} \frac{R}{a}$, which
is proportional to the driving strength. To zero order, one can assume
$a\simeq const$and the phase equation effectively decouples from the amplitude
equation, which in turn is driven by the phase, similar holds for a first
order \cite{Pikovsky-Rosenblum-Kurths-01}. The Adler equation
($q(\phi)=\sin(\phi)$) has two stationary solutions, $\dot{\phi}=0$, for
$|\Delta \omega| \leq \epsilon$. The stability of these fixed points, $\phi_s$
and $\phi_u$, is determined by linear stability analysis.  One puts
$\phi=\phi_{s/u}+\delta\phi$, inserts this expression into (\ref{eq:Adler})
and Taylor-expands the expressions. The stable point is given by $\cos \phi_s
< 0$, the unstable one by $\cos\phi_u>0$, corresponding to exponential decay
or growth of the perturbation $\delta \phi$.  In general, if higher harmonics
enter in $q(\phi)$ several fixed points can exist and the positions can be
strongly asymmetric, cf. Fig.~\ref{fig:Adler}.\\[-1.25ex]

%\subsubsection{The transition to synchronization}
\underline{\it The transition to synchronization:}
If one varies the detuning from zero to $\epsilon$ the phase varies from zero
to (at most) $\pi$. At the merging -, or bifurcation point, $\Delta
\omega_{max}-\epsilon=0$, and the stable and unstable fixed point annihilate (opaque
in Fig.~\ref{fig:Adler}). This type of bifurcation is known as saddle-node
bifurcation \cite{Ott-94}. In the case of an unstable saddle one attracting
and one repelling direction to/from the periodic limit cycle exist, for a
stable cycle two attracting directions exist, as for a stable node.
%Along the cycle, one finds marginal stability.

Inside the synchronization region, the phase difference $\phi = \phi_0=const$, and the
phase of the observed oscillation is $\omega_1 t + \phi_0$.  Let us turn to
the phase dynamics outside the synchronization region.  For $\dot{\phi}\neq
0$, one can formally integrate Eq.(\ref{eq:Adler}) to obtain
\begin{equation}
t= \left| \int_{\phi_0}^{\phi} \frac{d\phi^\prime}{\epsilon q(\phi^\prime) -
  \Delta\omega} \right|
\label{eq:Time}
\end{equation}
The period $T=2\pi/\omega_b$ results with the integration bound
$\phi=\phi_0+2\pi$. The total phase rotates uniformly $\varphi=\omega_1 t +
\phi(t) = \omega_1 t + \omega_b t = \Delta \Omega t$ and one recognizes the
typical beat phenomenon, as for the superposition of two harmonic waves; only
that here the beating is strongly nonlinear. For harmonic oscillations the phase-slip of the
beating is distributed over the whole interval. Close to the bifurcation from
the synchronized region, one observes  long epochs of nearly
constant phase $\phi\simeq \phi_{max}$ broken by relatively short intervals in
which the phase rotates by $2\pi$. We observe this behavior as well for
coupled organ pipes, this is illustrated below by a plot of the measured time
signal close to the synchronized region.

For a saddle-node bifurcation, one expects a square root dependence of $\Delta
\Omega$ on the detuning $\Delta \omega$. This indeed by expansion of the
denominator in (\ref{eq:Time}):
\begin{eqnarray}
\label{eq:SNBif}
|\Delta \Omega| &\simeq& 2 \pi \left|
\int_{-\infty}^{\infty}
\frac{d\phi}{\epsilon/2 q^{\prime\prime}(\phi_{max}) \phi^2 - (\Delta\omega -
  \Delta\omega_{max} )}
\right|^{-1}\;,\\
 &=& \sqrt{\epsilon |q^{\prime\prime}(\phi_{max})|\cdot(\Delta\omega -
  \Delta\omega_{max})} \simeq \sqrt{\Delta\omega -  \Delta\omega_{max}}
\end{eqnarray}
A very detailed description, including many examples and generalizations is
given in \cite{Pikovsky-Rosenblum-Kurths-01}.\\[-1.25ex]
%To summarize the procedure: one
%has to assume a limit cycle solution with small inharmonicity, apply the
%method of averaging over fast varying terms to obtain the equations for the
%amplitude and phase. In the phase equation, to first order, there appear two
%terms, the detuning and a periodic function which depends linearly on the
%coupling strength. This function can yield stable stationary solutions for the
%phase $\phi$, in dependence on the detuning and the coupling. The maximum
%span of the synchronization region is thus $\pi$. As long as such a solution
%exists, the system will be synchronized with the external force, if not it
%will find another, non-synchronized solution.

%\subsubsection{Two coupled oscillators}
\underline{\it Two coupled oscillators:}
Now, we will turn our attention to the case of two oscillators and focus to
the equations for the phases. From the above, it is now clear that any pair of
uncoupled, self-sustained oscillators close to a limit cycle can be written in terms of phase,
($\Theta_1,\Theta_2$), and amplitude, ($A_1,A_2$), in the following form:
\begin{eqnarray}
\dot{\Theta}_i &=&  \omega_{i} \\
\dot{A_i} &=& -\gamma\, (A_i-A_{i,0})\,,
\end{eqnarray}
with $i=1,2$. For weak coupling, one can again perform an expansion and apply
the method of slowly varying amplitude such that  the phase equations are,
analogous to (\ref{eq:Adler})
\begin{eqnarray}
\dot{\Theta}_1 &=&  \omega_{1} + \epsilon G_1(\Theta_1,\Theta_2)\;,\\
\dot{\Theta}_2 &=&  \omega_{2} + \epsilon G_2(\Theta_1,\Theta_2)\;.
\end{eqnarray}
For $\omega_1\simeq \omega_2$ the phase difference $\phi=\Theta_1-\Theta_2$ is a slow
variable. The discussion can now follow the above, with the difference  that
now the oscillators have to adjust themselves mutually, whereas above one
oscillator adjusted its frequency to an external driving.

We have to distinguish very carefully and clearly between the {\it single}
angular frequencies $\omega_i$, measured for a {\it decoupled} system, and the
angular frequencies $\dot{\Theta}_i$ measured for the {\it coupled}
system. For a clear nomenclature, we call the difference of the uncoupled
frequencies {\it detuning}, and denote it by the symbol $\Delta f =
(2\pi)^{-1} (\omega_2-\omega_1)$. The {\it frequency difference} of the
coupled system is always denoted by $\Delta \nu$. A typical plot visualizes
$\Delta \nu$ against the detuning $\Delta f$.  One observes a
plateau $\Delta \nu=0$ for a synchronized region and a linear dependence for
uncoupled oscillations.  Below, we will use the symbol $f$ for uncoupled
frequencies, and $\nu$ for measurements of the coupled pipes, analogously we
use $\Delta f$ and $\Delta \nu$.

At the end of this section, we would like to hint to two facts: i) the theory
presented above holds only for small deviations from harmonic oscillations and
for small coupling. For large parameters, in principal the considerations
still hold, but quantitatively deviations are expected.  ii) Even though it
seems appealing to explain the frequency locking of organ pipes in such a
simple way, it is not completely satisfactory because organ pipes are extended
systems and such a simple description does not take into account the
aeroacoustics which is eventually needed for a complete understanding.

\section{Measurement and Results}
\label{sec:experiment}

\subsection{Measurement Setup and Signal Analysis}

\begin{table}[b!]

%\begin{center}
 \caption{Pipe geometry.
 \label{tab:geometry}}
 \vspace{0.2cm}
\begin{tabular}{l r}
 \hline
  \multicolumn{2}{l}{\textbf{pipe-body}}\\
 \hline
 overall length (without foot)  & \unit[530]{mm}\\
 clear length (adjustable)      & $\unit[370\ldots 450]{mm}$\\
 wall thickness                 & $\unit[6]{mm}$\\
 \hline
  \multicolumn{2}{l}{\textbf{cross section}}\\
 \hline
  shape                         & quadratic\\
  clear width                         & \unit[37]{mm}\\
  clear depth                         & \unit[48]{mm}\\
 \hline
  \multicolumn{2}{l}{\textbf{mouth}}\\
 \hline
  cut-up height                 &\unit[7]{mm}\\
  cut-up width                   &\unit[37]{mm}\\
  flue-exit height               &\unit[0.4]{mm}\\
 \hline
  \multicolumn{2}{l}{\textbf{foot}}\\
 \hline
 length                         & \unit[320]{mm}\\
 toe hole diameter              & \unit[15]{mm}\\
 \hline
\end{tabular}
%\end{center}
\end{table}

Measurements were carried out on a miniature organ especially made
by Alexander Schuke GmbH  for physical investigations.
Just like a real organ it consists of a blower connected by a
mechanical regulating-valve to the wind-belt and finally a
wind-chest upon which the pipes are positioned. Since only
stationary behavior was of interest we joined the two pipes directly
to the wind-belt with flexible tubes. The wind pressure was set by
the organ builder to a mean value of \unit[160]{Pa}. Over
\unit[10]{s} we measured a standard deviation value of \unit[6]{Pa}
in the pipe foot. The wind may be considered as very stable.

The pipes used are stopped, a detailed description of the pipe-geometry is given
in table \ref{tab:geometry}. The resonator top of the pipes was
movable in order to adjust resonator-length and pitch. To minimize
losses, a gasket made of felt was applied. The pipes as delivered
from the manufacturer were tuned to e and f in German notation, that
is E3 and F3 in American notation. The applied driving pressure
roughly yields a Reynolds number $Re=435$.

The acoustic signal was measured by a B\&K 4191 condenser microphone
positioned at the centerline between the sound sources. Further information on
positioning of the acoustic sources and the microphone is given in sections
\ref{sec:acoustic_coupl} and \ref{sec:hydrodyn_coupl}.  Measurements took
place inside the anechoic chamber of the Technical University of Berlin. The
rooms size is $W \cdot L \cdot D = \unit[20]{m} \cdot \unit[17]{m} \cdot
\unit[7]{m}$ giving a volume of $\unit[2000]{m^3}$. The low cut-off frequency
of this room is \unit[63]{Hz}. Temperature, humidity and pressure conditions
were kept stable to keep there influence on the measurements negligible.

\begin{figure}[h]
\includegraphics[width=0.7\textwidth]{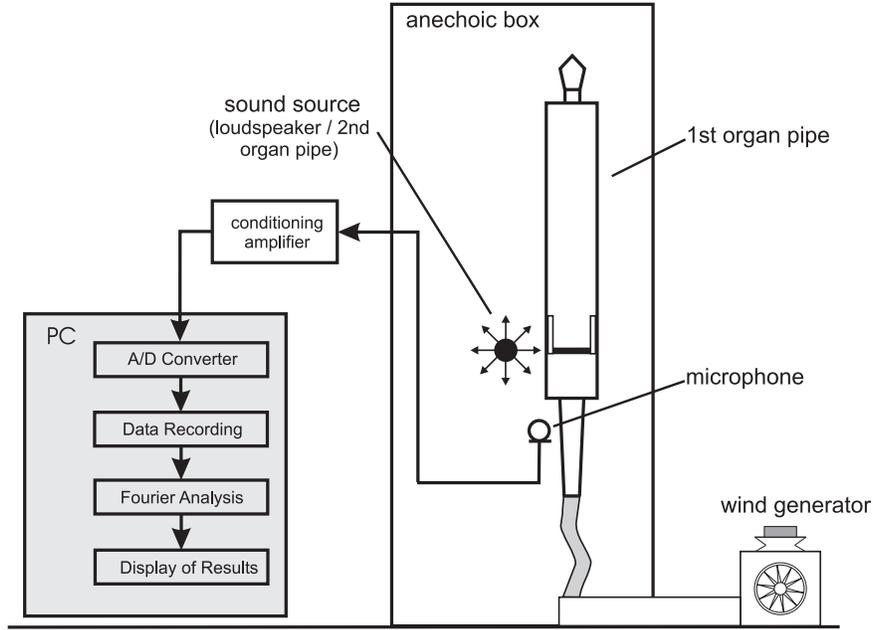}
\caption{Sketch of the experimental setup. The microphone is in plane
  with the pipe mouth, the view is at an angle from above.}
\label{fig:Setup}
\end{figure}

%\subsection{Signal Analysis Procedure}
The recording of the sensor signal as a function of time was achieved via
hard-disk recording with a M-Audio 2496 Soundboard at a sampling rate of
\unit[44.1]{kHz} and a resolution of \unit[16]{bit} and saved in PCM
format. Data analysis was performed with $\textsc{Matlab}$. All Spectra shown
in the results are Fourier transformed out of \unit[10]{s}
(\unit[441\hspace{.3ex}000]{samples}) of time signal, giving a
frequency-resolution of \unit[0.1]{Hz}. During Fourier transformation a
symmetric Hann window was applied to time data. All spectra have been
averaged.  From these spectra the amplitude of the harmonics was read out by a
routine searching for the maximum signal level within a frequency range
of~$\unit[\pm 0.3]{Hz}$.

%\section{Results and Physical Modeling}

\subsection{Organ Pipe driven externally by a Loudspeaker}
\label{sec:acoustic_coupl}

\begin{figure}[h]
  \includegraphics[width=0.44\textwidth]{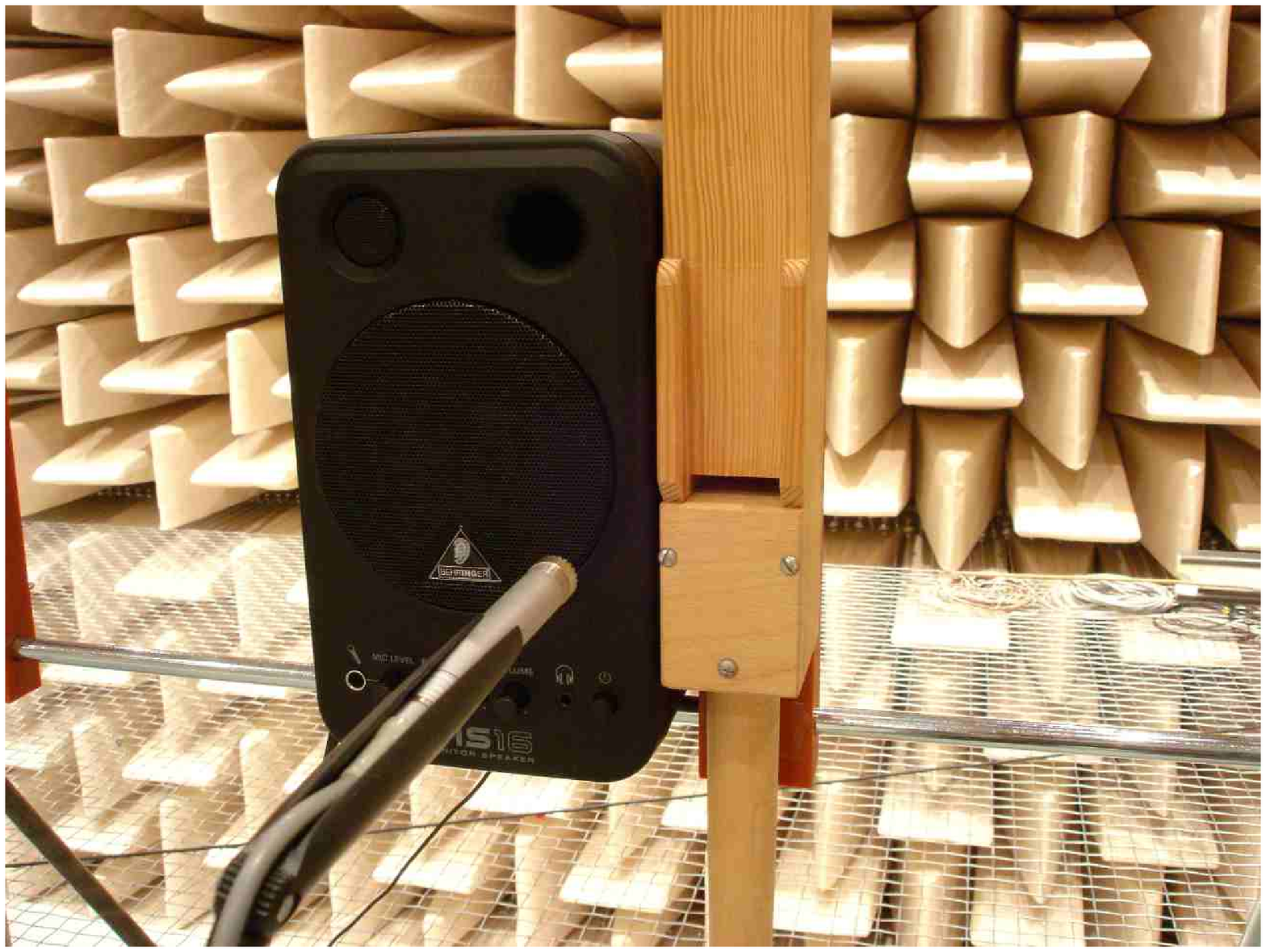}
  \hspace{2ex}
  \includegraphics[width=0.44\textwidth]{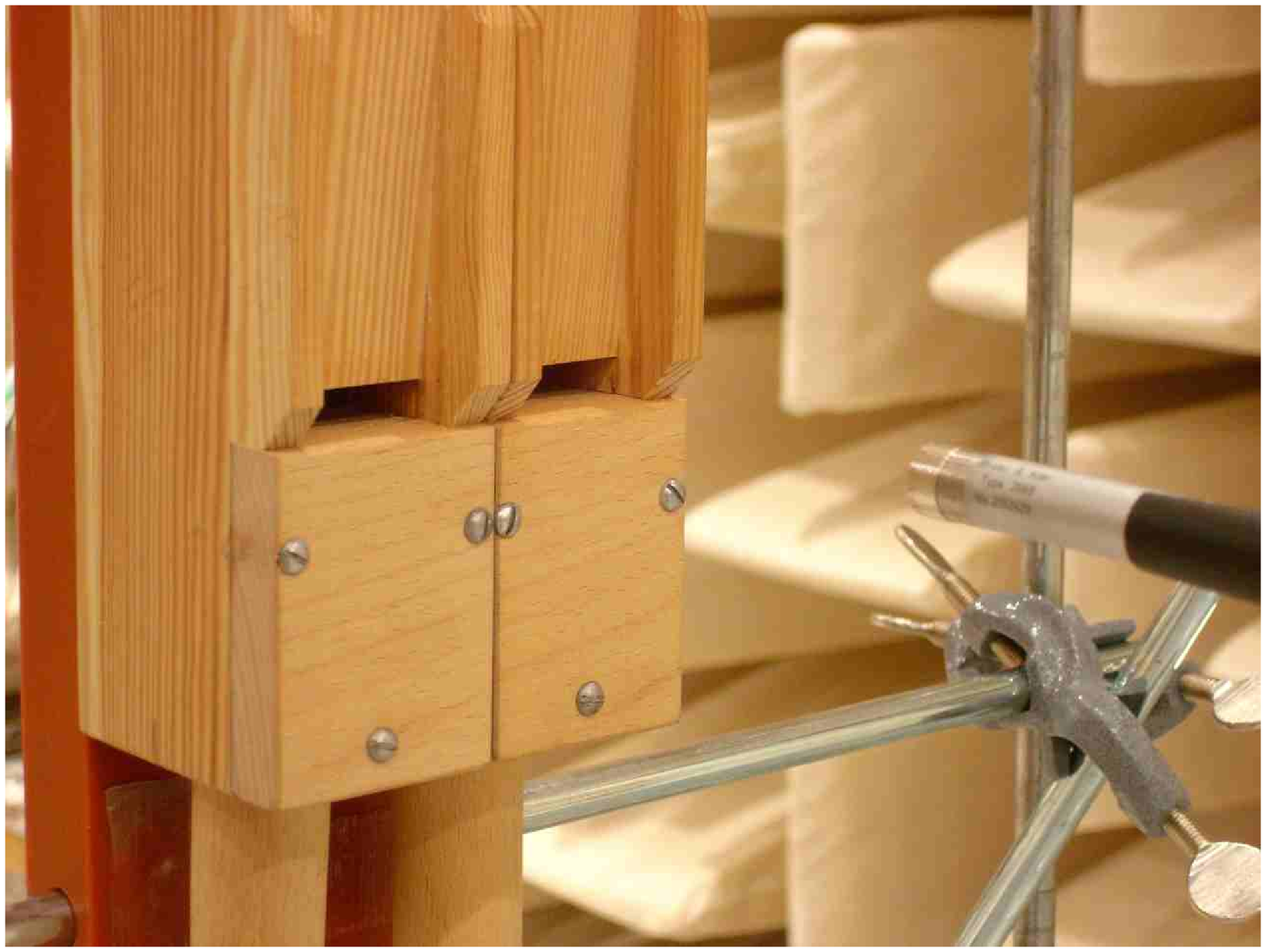}
  \caption{Fotograph of the measurement setup. 
    Left: loudspeaker and pipe sounding
    together, right: two ``coupled'' organ pipes }
\label{fig:SetupFoto}
\end{figure}

To investigate the influences of a external sound source onto the organ-pipe
we positioned ax loudspeaker directly beside a single organ-pipe.  The
loudspeaker was a 2-Way Active System MS 16 from Behringer. The woofers
cone-diameter is \unit[10]{cm}. The speaker was connected to a HP 33120A
waveform generator, delivering a sinusoidal signal. The sound pressure level
emitted by the loudspeaker was set equivalent to the volume of the fundamental
organ tone within $81.5\pm \unit[0.25]{dB}$. At this level the distortion of
the speaker was at THD=1.5\%. The experiment started with the pipe at a
fundamental frequency of \unit[168.3]{Hz} and the loudspeaker at
\unit[165.8]{Hz}. During the measurement the organ-pipe remained unchanged but
the frequency of the loudspeaker was very slowly, every 40 seconds, increased
by \unit[0.1]{Hz}. This gives the possibility to average ten seconds of
signal four times to suppress noise. The microphone was positioned at the
centerline between pipe and loudspeaker at a distance of \unit[15]{cm}. A
picture of the setup is given in Fig.~\ref{fig:SetupFoto}.

\begin{figure}[h]
\includegraphics[angle=0,width=0.8\textwidth]{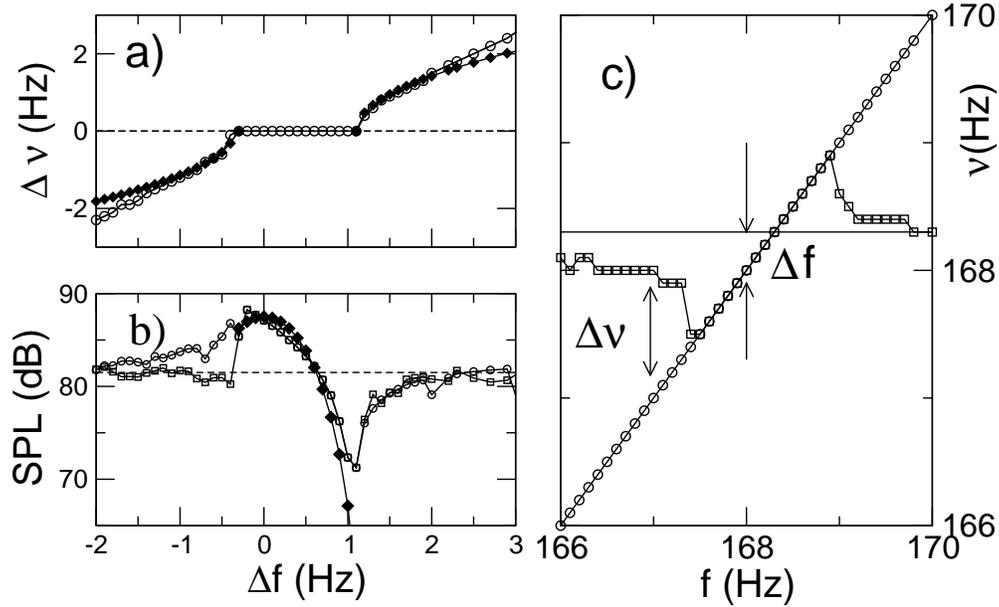}
  \caption{Upper left (a): Frequency locking of organ pipe and
    loudspeaker. Top: observed frequency difference $\Delta \nu$ ascertained
    with pipe and loudspeaker sounding together versus detuning $\Delta f$
    (open circles). In the synchronization region, a very clean plateau is
    observed. A saddle-node bifurcation (filled diamonds) occurs at the edges,
    as predicted by Eq.~(\ref{eq:SNBif}).  Bottom left (b): Measured amplitude
    signal in dB. The phase varies from $0$ to $\pi$ over the synchronization
    plateau. The region is asymmetric; $\Delta \nu \in (-0.3,1.1)$ within the
    frequency resolution.  The circles indicate the peak amplitude at the
    loudspeakers frequency, and the pipes frequency (squares). The
    superposition of the signals including the phase shift along the plateau
    is plotted by filled diamonds, the agreement is quite good.  Right (c):
    absolute measured signals for pipe and loudspeaker sounding together
    ($\nu$, open circles) and the signals for speaker alone (squares) and pipe
    alone (straight line). The detuning is indicated by $\Delta $f, as the
    difference of the ``uncoupled oscillators'', pipe and speaker alone; the
    frequency difference, $\Delta \nu$ results from the subtraction of the
    measured signal for pipe and speaker sounding simultaneously, coupled and
    the varied frequency, in our case of the loudspeaker. }
\label{fig:PipeSpeaker}
\end{figure}

The synchronization can be observed clearly by the plateau in the plot of
frequency difference, $\Delta \nu$ against detuning, $\Delta f$. As well the
predicted square root function is observed at the transition to
synchronization, cf. Fig.~\ref{fig:PipeSpeaker}. The amplitude recorded with the
microphone shows a positive and negative peak, at the beginning and end of the
synchronization, respectively. Invoking the explanations of synchronization
theory, cf. Eq.~\ref{eq:Adler}, this amounts to a change of the relative phase
from 0 to $\pi$. If loudspeaker and pipe are considered as monopole sources,
their field at the centerline, where the signal is recorded, can be obtained
as a superposition of two fields with equal, measured amplitude and phase
difference $\phi_0$, decaying as $1/r$: $A(r)=A_0/r+A_0/{r}\cos\phi_0$
(remember that we adjusted the sound pressure level of loudspeaker and pipe,
so $A_{speaker}=A_{pipe}=A_0$).  For $\phi_0=0$
interference is positive, $\phi_0=\pi$ implies negative interference. This is
exactly seen in Fig.~\ref{fig:PipeSpeaker}b, where we plot the peak amplitude
of the loudspeakers frequency (open circle), and the first harmonic of the
organ pipe (filled circles), respectively, over the detuning. For zero
detuning, $\phi_0=\pi/2$.  This corresponds to an interval of $\pi$ within the
synchronization region; i.e. from positive to negative superposition.  One
notices that the phase $\phi_0=\pi/2$ right at $A=A_0$. This corresponds to an
equal distribution of energy at this point. The increase of the amplitude at
the loudspeakers frequency at the left hand of the synchronization region
comes from the above mentioned effect that the pipe sounds for long periods
with the frequency of the speaker to undergo a fast phase slip, when close to
be synchronized. This results in a higher peak amplitude at the loudspeakers
frequency (cf. the discussion in Sec.~\ref{sec:Coupling}).

\subsection{Two Pipes standing side by side}
\label{sec:hydrodyn_coupl}

To measure the dependence of frequency locking on the detuning of
the two pipes, both pipes were installed on a common bar directly
side by side (see Fig.~\ref{fig:SetupFoto}.
One of the pipes has been kept at the fixed frequency of $170.1 Hz$,
the other pipe was detuned in variable steps. With two pipes
connected to the wind-chest one cannot detect the fundamental
frequency of one pipe by simply turning of the other through closing
the pipes valve. This would raise the wind pressure for the
investigated pipe and consequently also the frequency itself. In
order to acquire the value precisely the pipe at a fixed frequency
was made soundless by putting a little absorber-wedge into the
mouth. Doing so the air--sheet oscillation and therefore sound
generation is suppressed, but the air consumption remains the same,
keeping the wind-pressure stable.

\begin{figure}[h]
\includegraphics[width=0.8\textwidth]{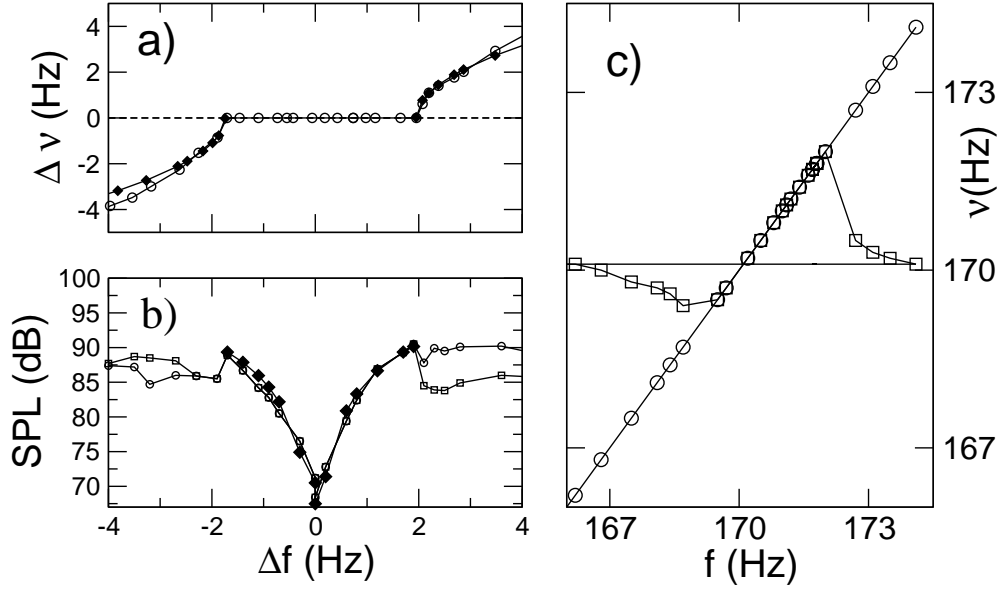}
\caption{Frequency locking of two organ pipes. The plot shows the observed
frequency difference $\Delta \nu$ versus the detuning of the uncoupled pipes,
$\Delta f$ (circles). Upper left (a): In the synchronization region, a very
clean plateau is observed.  As in the case of external driving, the
saddle-node bifurcation is observed at the edge to the synchronization region
(filled diamonds). The agreement is very good.  Lower left (b): Amplitudes at
the first harmonics of the pipes (circles, variable-frequency pipe;squares,
fixed-frequency pipe) against detuning; the sharp decrease of the amplitude at
$\Delta f=0$ indicates an antiphase oscillation at the pipe mouth. The
analytical curve obtained using synchronization theory is shown with filled
diamonds, assuming $\phi_0=\pi$---the agreement is excellent.  Right (c):Plot
of the absolute frequencies (circles/squares as above),, the straight line
shows the \unit[170.1]{Hz} of the fixed-frequency pipe.}
\label{fig:PipePipe}
\end{figure}

In \cite{Stanzial-00}, the frequencies $\nu_{1}$, $\nu_{2}$
of the coupled pipes were plotted against the frequency detuning of
the uncoupled pipes, $\Delta f$. As explained above, a clearer
characterization is achieved when the frequency difference $\Delta
\nu$ is plotted versus $\Delta f$
\cite{Bogoliubov-Mitropolsky-61,Pikovsky-Rosenblum-Kurths-01}. We
plot this quantity in Fig.~\ref{fig:PipePipe}. Clearly, the typical
behavior of frequency locking can be observed. As synchronization
theory predicts, the bifurcation close to the ends of the locking
region is of saddle node type \cite{Ott-94} (filled diamonds in
Fig.~\ref{fig:PipePipe}a).

Now, stressing the analogy with two coupled oscillators, a change of
the relative phase from $\phi_0-\pi/2$ to $\phi_0+\pi/2$ is expected
\cite{Pikovsky-Rosenblum-Kurths-01} ($\phi=2\pi f$).  This has
consequences for the amplitude of the emitted sound. The radiated
sound wave in the far field can be roughly approximated by a
spherical wave, and the field at the measurement point $\vec{r}$ is
obtained by superposition of the two waves emitted at the pipe
mouths positions. Assuming the sources to be point-like and
considering that the pipes are situated directly side-by-side, the
amplitude along the center line is $A={2A_0}/{r}\,(1+\cos\phi_0)$,
this has already been described in \cite{Bouasse-29}. The notation
is as above, please note that here $A_{pipe 1}=A_{pipe 2}=A_0$,
because both pipes have approximately the same amplitude, cf.
Fig.~\ref{fig:PipePipe}.

\begin{figure}[b!]
\includegraphics[clip=true,width=0.9\textwidth]{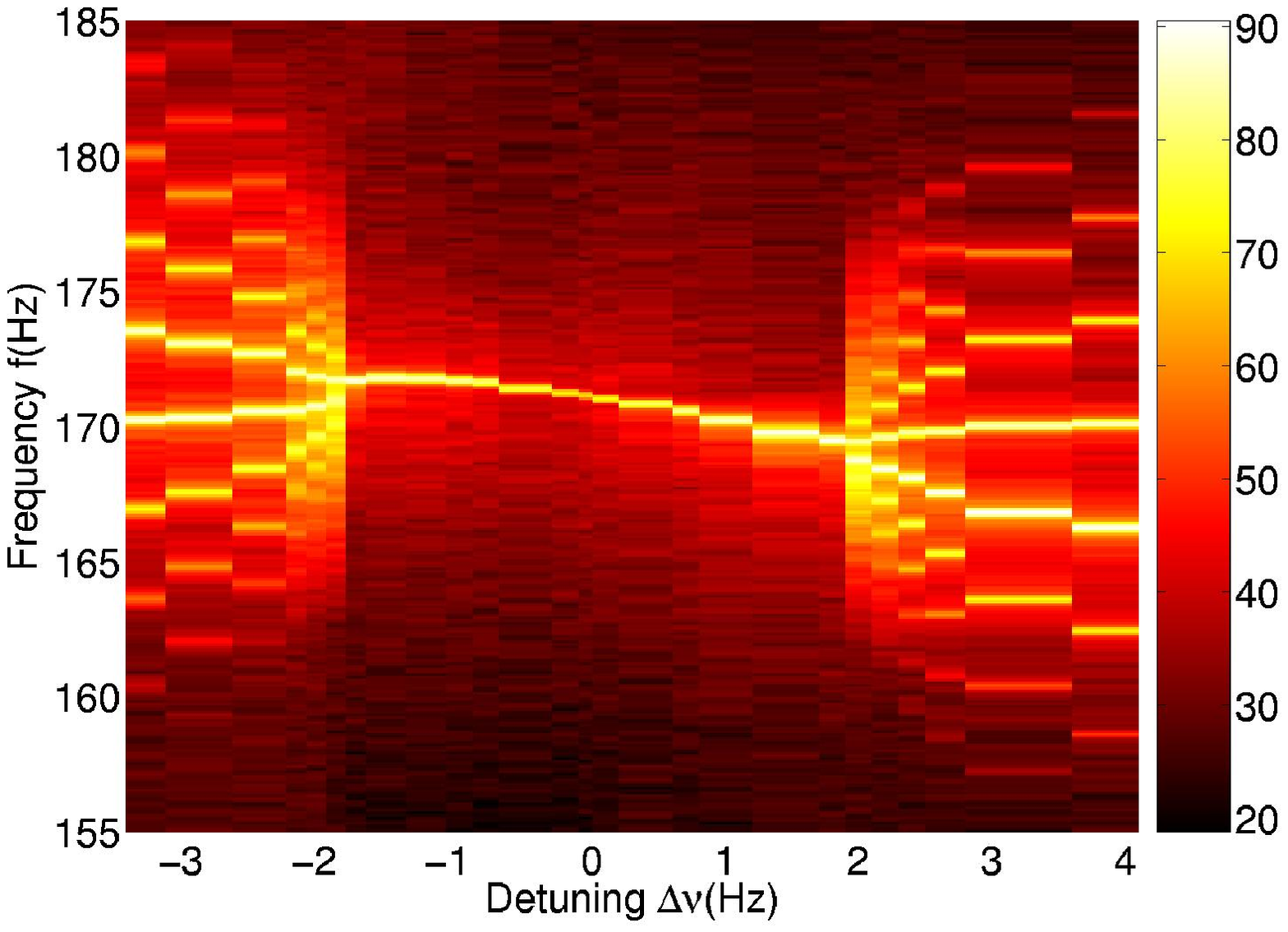}
\caption{A sharp transition to synchronization is observed and the sidebands
  from beating in the measured spectrum collapse to a single frequency in the
  synchronization region. The amplitude is encoded according to the levels on
  the right.}
\label{fig:Locking2}
\end{figure}

Varying the frequency mismatch $\Delta f$ we observe a gradual increase of the
amplitude at the centerpoint, where the microphone is situated. In
Fig.~\ref{fig:PipePipe}b, the result of the measurement is displayed: We see
an excellent agreement of our estimate with the measurement.  This behavior is
observed as well in the spectral plot, Fig.~\ref{fig:Locking2}.  where one can
observe the collapse of all sidebands in the measured spectrum to one single
frequency as a very sharp transition.

\begin{figure}[b!]
\includegraphics[width=0.7\textwidth]{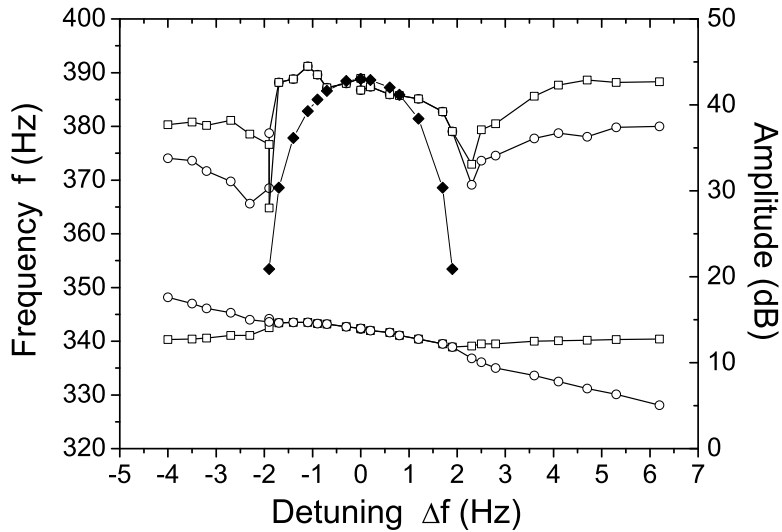}
\caption{Amplitude and frequency difference for the second harmonic. Here, the
  phase is doubled and an annihilation occurs right at the edge of the
  synchronization region, seen as a sharp decrease. The curve predicted by
  theory shows deviations for the second harmonic.}
\label{fig:Amplitude2}
\end{figure}

For higher harmonics, the phase difference by $\phi_0$ does not imply
destructive interference because they are slaved and follow the first
harmonic. In other words, the relative phase for the $n$th harmonic in the
synchronization region lies in the interval
($n(\phi_0-\pi/2),n(\phi_0+\pi/2)$). For the second harmonic, this implies
in-phase oscillations at $\Delta f=0$.  This is heard acoustically by a
dominance of the octave around $\Delta f=0$. At the edge of the
synchronization region, however, there is a phase difference of $\pi$ and
destructive interference is expected. This is confirmed well by the
quantitative measurement of the amplitude dependence of the second harmonic,
shown in Fig.~\ref{fig:Amplitude2}. The agreement with the estimate from
superposition of two monopoles is still good, but deviations occur. Higher
harmonics are in qualitative accordance with the above ideas, but
quantitatively more and more differences are observed. This can be expected,
because the approximation by a monopole source does not hold.

When two pipes are coupled diffusively, a so-called oscillation death can
occur, leading to a complete silence of both pipes. The whole input power
would then be converted to heat without radiation.  We thus observe a strong
impact of the synchronization of the pipes on the acoustic properties of the
emitted sound. From our results, we can rule out an oscillation death for two
coupled pipes with zero detuning, as suggested in
\cite{Pikovsky-Rosenblum-Kurths-01}. Rather, the two oscillators radiate two
antiphase waves; this results in a vanishing acoustic signal.

\subsection{The Coupling of the Pipes}
\label{sec:Coupling}

As in the case of the external driving by a loudspeaker, the two pipes
interact by the pressure perturbation generated at the pipes mouths. But now, there are
two sources of pressure contributions: the acoustic and the aerodynamic part
from the vortices itself. Which one is dominant depends on the ratio of their
amplitudes. In the case of the organ pipes, the acoustic pressure is very
large due to the selected amplification of the resonator frequencies. However,
the wind field at one pipe can literally ``blow away'' an adjacent vortex. On
the basis of our observations, we can only state that definitely, the
acoustical field is sufficient to give raise of synchronization. The influence
of the wind field must be subject of careful future work including flow
visualization on the  scale of several vortex diameters.
At the edge of the synchronization region, there must be a phase difference of
$\phi_0 \pm\pi/2$, which corresponds to a shift of a quarter of a jet
oscillation.  The jets then undergo a mutual reordering into a complex
spatio-temporal three dimensional pattern.  Again, the observation of the
pattern was beyond the facilities of the current experimental setup.

\begin{figure}[b!]
\includegraphics[angle=0,width=0.7\textwidth]{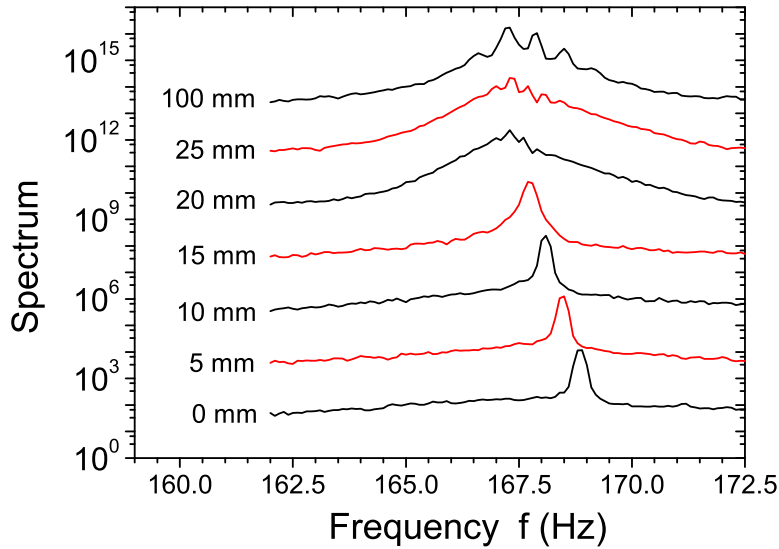}
\caption{Frequency spectrum of the coupled pipes in dependence on the
  distance, given in the legend. The pipes decouple more and more when farther
  apart. For small distances, a sharp peak at the synchronization frequency is
  observed, for large distances, the peak broadens more and more, at very
  large distances the spectrum of the uncoupled pipes is recovered with the
  typical beat phenomenon. To distinguish different graphs, an offset has been
  added to each curve.}
\label{fig:CouplingPipe}
\end{figure}

All our observations are consistent with an interpretation in the
frame of synchronization theory of two oscillators. A full
description of the physics has to take into account the origin of
the sound generation---the oscillations of resonator and the
vortices generated by two jets. Most likely, there is a competition
of the interaction between jets and resonators. Which mechanism wins
depends then on the ratio of the respective amplitudes. A comment
should be made on why the synchronization description by two
oscillators holds so well. In the measured Reynolds number regime,
there might be three-dimensional patterns at the pipe mouth, if the
aspect ratio is large enough. To check this in our particular case
requires detailed experiments.  Neglecting three dimensional
structures, and assuming that the air sheet is homogeneously
oscillating, a description by oscillator models like Stuart-Landau
or a van-der-Pol equation could be sufficient to describe the two
dimensional oscillations and the transition to turbulence
\cite{Provansal-87}. Since the main sound production is due to the
interaction with the labium, an oscillator model suits well. From
synchronization theory it follows that {\it any} self-sustained
oscillators will generically follow the synchronization scenario
when coupled, leading to the observed behavior
\cite{Pikovsky-Rosenblum-Kurths-01,Aronson-Ermentrout-Koppel-90}.
The simulations in \cite{Stanzial-00} are completely in agreement
with this fact. We are convinced that full understanding may be
achieved by studying the details of the coupling mechanism.

To investigate the sensitivity to coupling strength, we measured the frequency
spectrum about the first harmonic while varying inter-pipe distance (measured
between the outer walls of the pipes).  The detuning was fixed to \unit[0.7]{Hz}.
The corresponding plot is depicted in Fig.~\ref{fig:CouplingPipe}. For large
distances, the typical interference pattern of two noninteracting oscillators
is observed.  As the pipes come closer to each other, the individual, sharp
peaks of half width $\simeq$ 0.1 Hz reduce in amplitude and at the same time
the spectrum broadens to a ``hump'' with half width $\simeq 1$ Hz. Peaks
typical of the beating from linear superposition sit on the hump. Coming even
closer, full synchronization is observed with one single, very sharp peak of
again half width $\simeq 0.1$ Hz. That means the hump is about ten times
broader than either of the peaks for the uncoupled system or the synchronized
one.

\begin{figure}[b!]
\includegraphics[angle=0,width=0.4\textwidth]{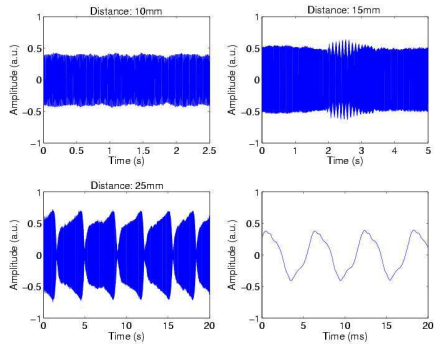}
\caption{Time signal for three different distances. Upper left: 10mm, lower
  left: 15 mm, upper right:25 mm. The lower right graph shows a typical fast
  oscillation as found for all three distances. The transition from
  synchronization to beat happens between 15 and 25 mm. For 10 mm the
  amplitude is constant and the phase rotates with the fast frequency. At 25
  mm the beat phenomenon is observed with a constant amplitude over relatively
  long times (e.g. ) and a quick phase slip (). At 15 mm, the system shows
  some intermittent bursts at irregular times. }
\label{fig:HumpTime}
\end{figure}

The basic observations can be explained by synchronization theory. In
Eq.~\ref{eq:SNBif}, the square root behavior of the frequency difference at
the transition to synchronization (the bifurcation point) is derived. Lowering
the coupling strength is equivalent to shift the bifurcation point towards
zero. At a certain distance the detuning leaves the synchronization region and
one should observe a (strongly nonlinear) beat. The peaks of the beat are
observable as small side peaks on top of the broad peak. For a pure beat, only
sharp peaks should be observed. Here, the duration of the large time interval
-we do not want to call it period- varies slightly such that the sidebands
wiggle a bit, altogether this leads to a broad peak. To
illustrate this behavior, we plot in Fig.~\ref{fig:HumpTime} three different
time signals for the distances \unit[10]{mm} (synchronized), \unit[15]{mm}
(shortly before synchronization), and \unit[25]{mm}. The fast oscillations
look very similar for all of the distances, as exemplary given by
Fig.~\ref{fig:HumpTime} (bottom, right).

A complete understanding why such a broad hump appears could not be achieved.
There are several possible explanations. According to classical
synchronization theory the frequency fluctuation cannot be explained but by
noise, e.g.  in the air supply in the wind system.  Another option are the
small fluctuations in the vortices at the pipe mouth which can be sufficient
to cause the very small variation in the beat frequency. Finally, one can
consider the bifurcation to synchronization as a kind of phase transition,
where typically large fluctuations occur, including chaotic behavior. By means
of our measurement we cannot decide which of these scenarios is true. Finally,
we want to mention recent results concerning whistling of two nearby Helmholtz
resonators show a similar transition in the phase of the acoustical signals
emitted by the resonators \cite{Derks-Hirschberg-04}. Interestingly, the phase
relation is lost at distances of the order of the ones we find (25mm, if the
wall thickness of the pipes is included). Thus, this observation might be
explained by synchronization as well.

\section{Conclusion}
\label{sec:Conclusion}

We presented measurements on an external driving of an organ pipe by a
loudspeaker and on the mutual influence of two organ pipes.  The observed
behavior is completely consistent with an explanation in the frame of
synchronization theory. For the dependence of the synchronization on the
``coupling strength'', measurements determining the dependence of the
frequencies on the inter-pipe distance have been carried out. They reveal a
broadening of the peak at the synchronization frequency with increasing
distance. This broadening can be explained by quick phase slips as predicted
by theory close to a saddle-node bifurcation. The smearing out of the side peaks
of this beat can be either due to inconstant wind supply or due to the
fluctuations caused by the oscillating air sheet at the pipe mouth. To clear
this question further measurements have to be carried out.

From an acoustic point of view, one can address the question of how
to position two organ pipes close in frequency. This has been
intuitively solved by organ builders by trial and error in the last
centuries \cite{Angster-93}. Our work might give quantitative hints
on how large the inter-pipe distance needs to be to suppress mutual
influence, and on details of the coupling mechanism.  For example,
avoiding an amplitude minimum for the first harmonic is highly
desirable.

From an aerodynamical point of view, the above scenario requires more detailed
investigations to understand the full dynamics of the coupled resonator/jet
system, although a model consisting of two mutually coupled oscillators seems
to be sufficient for all qualitative questions. In addition the more involved
setup of more than two organ pipes is an interesting subject for further
investigations.  From our measurements, one can clearly say that the pipes do
not show an oscillation death; rather, antiphase sound radiation yields the
observed weakening of the amplitude.

\section*{Acknowledgements}
We acknowledge fruitful discussion with M. Rosenblum and A. Pikovsky about
synchronization theory, with D. Lohse and J. F. Pinton, and thank J. Ong for
careful reading of the manuscript. M. Abel acknowledges support by the DFG
(Deutsche Forschungsgemeinschaft). M. Abel and S. Bergweiler are supported by
UP Transfer Potsdam. We thank the organ manufacturer Alexander
Schuke Potsdam Orgelbau GmBH for kindly providing the organ pipes for our
measurements.

\end{document}